*Research Article*

# IoT-Based Remote Health Monitoring System Employing Smart Sensors for Asthma Patients during COVID-19 Pandemic

Nafisa Shamim Rafa,[1] Basma Binte Azmal,[1] Abdur Rab Dhruba,[1] Mohammad Monirujjaman Khan,[1] Turki M. Alanazi,[2] Faris A. Almalki,[3] and Othman AlOmeir[4]

[1]*Department of Electrical and Computer Engineering, North South University, Bashundhara, Dhaka 1229, Bangladesh*
[2]*Department of Electrical Engineering, College of Engineering, Jouf University, Sakaka, Saudi Arabia*
[3]*Department of Computer Engineering, College of Computers and Information Technology, Taif University, P.O. Box 11099, Taif 21944, Saudi Arabia*
[4]*Department of Pharmacy Practice, College of Pharmacy, Shaqra University, Shaqra, Saudi Arabia*

Correspondence should be addressed to Mohammad Monirujjaman Khan; monirujjaman.khan@northsouth.edu

COVID-19 and asthma are respiratory diseases that can be life-threatening in uncontrolled circumstances and require continuous monitoring. A poverty-stricken South Asian country like Bangladesh has been bearing the brunt of the COVID-19 pandemic since its beginning. The majority of the country's population resides in rural areas, where proper healthcare is difficult to access. This emphasizes the necessity of telemedicine, implementing the concept of the Internet of Things (IoT), which is still under development in Bangladesh. This paper demonstrates how the current challenges in the healthcare system are resolvable through the design of a remote health and environment monitoring system, specifically for asthma patients who are at an increased risk of COVID-19. Since on-time treatment is essential, this system will allow doctors and medical staff to receive patient information in real time and deliver their services immediately to the patient regardless of their location. The proposed system consists of various sensors collecting heart rate, body temperature, ambient temperature, humidity, and air quality data and processing them through the Arduino Microcontroller. It is integrated with a mobile application. All this data is sent to the mobile application via a Bluetooth module and updated every few seconds so that the medical staff can instantly track patients' conditions and emergencies. The developed prototype is portable and easily usable by anyone. The system has been applied to five people of different ages and medical histories over a particular period. Upon analyzing all their data, it became clear which participants were particularly vulnerable to health deterioration and needed constant observation. Through this research, awareness about asthmatic symptoms will improve and help prevent their severity through effective treatment anytime, anywhere.

## 1. Introduction

The current coronavirus disease (COVID-19) pandemic is highly distressing, as the second wave appears to be far more hazardous than the first. In the second wave of severe acute respiratory syndrome coronavirus 2, India was one of the most affected countries (SARS-CoV-2). On September 15, 2021, the total number of infected people in India was 33,339,375, which is increasing rapidly [1]. Because of their close geographical proximity, this is concerning for Bangladesh because the Indian variety of SARS-CoV-2 is more hazardous than the other variants. COVID-19, a highly contagious viral disease produced by SARS-CoV-2, has wreaked havoc on the world's demography, killing over 2.9 million people worldwide, making it the deadliest global health outbreak since the 1918 influenza pandemic. People older than 60 years and those with medical problems are, in general, at a higher risk of being infected by SARS-CoV-2.

According to the World Health Organization estimates, there are approximately 226,906,513 COVID-19 cases worldwide [2]. When patients are affected, doctors suggest using an oxygen meter to monitor their oxygen levels so that any irregularities can be recognized and treated early.

Asthma is a long-term chronic respiratory disease. It is a lung condition that can affect people of all ages. It is one of the most common medical conditions. Wheezing, breathlessness, chest tightness, and coughing are all symptoms of asthma. It is mainly caused by allergens like dust or pollen and infections like colds and flu [3]. COVID-19 is a major concern for asthmatics, as they have a significant risk of catching the virus due to their weakened lung function and immune system [4].

Bangladesh is a densely populated country, making it difficult to find a doctor who can provide adequate therapy [5]. According to the Bangladesh Health Facility Survey from 2017, more than 70% of rural healthcare facilities lack essential requirements [6]. There are extremely few specialized doctors accessible to the large population. Receiving effective therapy is difficult because there are few treatment alternatives due to the lack of specialized hospitals equipped for coronavirus in many areas. It is challenging to acquire competent treatment in an emergency, which often results in fatalities.

A smart object is a physical object with an embedded processor, data storage system, sensor system, and network technology [7]. The demand for this technology is increasing in various sectors, transforming industrial segments like manufacturing, construction, and power distribution. The term "Internet of Medical Things" (IoMT) is used to describe the application of IoT technology in healthcare through the network of connected sensors that detect vital data in real time [8]. Today, there are about 4 million devices used in the healthcare sector to collect and process patient data in real time [9]). As an e-healthcare technology, the Internet of Things (IoT) is a dependable technology that we may employ to strengthen our emergency health monitoring system [10]. IoT principles are employed to connect accessible medical resources and provide patients with smart, dependable, and effective healthcare.

Collecting and analyzing so much human and non-human-related data is a cumbersome process that can be made easier by integrating IoT and cloud computing. The traditional human-reported surveillance system is inconvenient in providing valuable and immediate information about diseases, risk factors, and environmental conditions among large-scale populations, particularly during a global pandemic [11]. The COVID-19 pandemic has elevated the need for applying IoMT technology since the situation calls for treating patients while maintaining social distance. The elderly, bed-ridden patients or patients who are apprehensive about leaving their homes can be treated remotely. The medical staff can monitor their progress in real time, and emergency situations can be taken care of conveniently. Such remote treatment also reduces the strain on hospital beds and the overall healthcare system.

The objective of this research is to develop an IoT-based remote health monitoring system for asthma patients who are particularly vulnerable to COVID-19. This system monitors five parameters, heart rate, body temperature, room temperature, humidity, and air quality, while the patients are rested over various intervals of time. All the data is displayed in real time at the medical team's end through a mobile application, so the doctors, or even the relatives of the patients, can understand the results and act quickly in case the patient's conditions deteriorate.

*1.1. Related Work.* A similar IoT-based health monitoring system has been developed by Tamilselvi et al. in the study [12], which focuses on monitoring the conditions of comma sufferers in real time. Their research implemented smart sensors like temperature, heartbeat, eyeblink, and SPO2 (peripheral capillary oxygen saturation) sensors for fetching the patient's body temperature, coronary heart rate, eye movement, and oxygen saturation percentage. The data from the sensors is processed through the Arduino Microcontroller. They also used the concept of cloud computing, via which the patient's data is transmitted to the application of the concerned organization. In [13], the authors proposed a low-cost remote health monitoring system for the purpose of tracking heart rate and body temperature in the form of a wristband. The sensor data is sent to a low-power microcontroller MSP430, which then transmits the received data through a low-cost transceiver module nRF24L01. Their system also uses Raspberry Pi 3 microcontrollers because of its WI-FI protocol, which is necessary for communication with the medical team. One major drawback of this system is that there is no user interface to display collected data. In [14], the researchers designed an Arduino-based health monitoring system specifically for COVID-19 patients in quarantine. The prototype of this system comes in the form of a wearable bracelet that collects the temperature, heart rate, oxygen saturation, and location data through the GPS sensor in real time. It also has an integrated cough detection system using an AI model. Both doctors and patients are able to access data with the help of a synchronized API and an Android application. One slightly similar kind of model was proposed by the authors in [15], where a continuous monitoring system focuses on elderly patients who are prone to suffering from heart attacks. In this system, heartbeat and temperature sensors were used, connected to an Arduino Uno, which in turn was interfaced with an LCD display and Wi-Fi module to send the data to the IoT platform. The system sends an alert in case any abnormality in a heartbeat is detected. Another related study was developed in [16], where the authors developed a system for elderly patients to monitor their body temperature, heart rate, and galvanic skin response. Apart from the sensors, their system used both Arduino and Raspberry Pi microcontrollers for data sensing, processing, and transmitting in real time. An android application was built to display the graphical representation of the data stored in the cloud. In [17], the authors proposed a system that can monitor the patients and it will show the data to a website. The authors use several health-related sensors similar to those in this study. But the use of websites instead of mobile applications reduces the mobility and flexibility of the system.

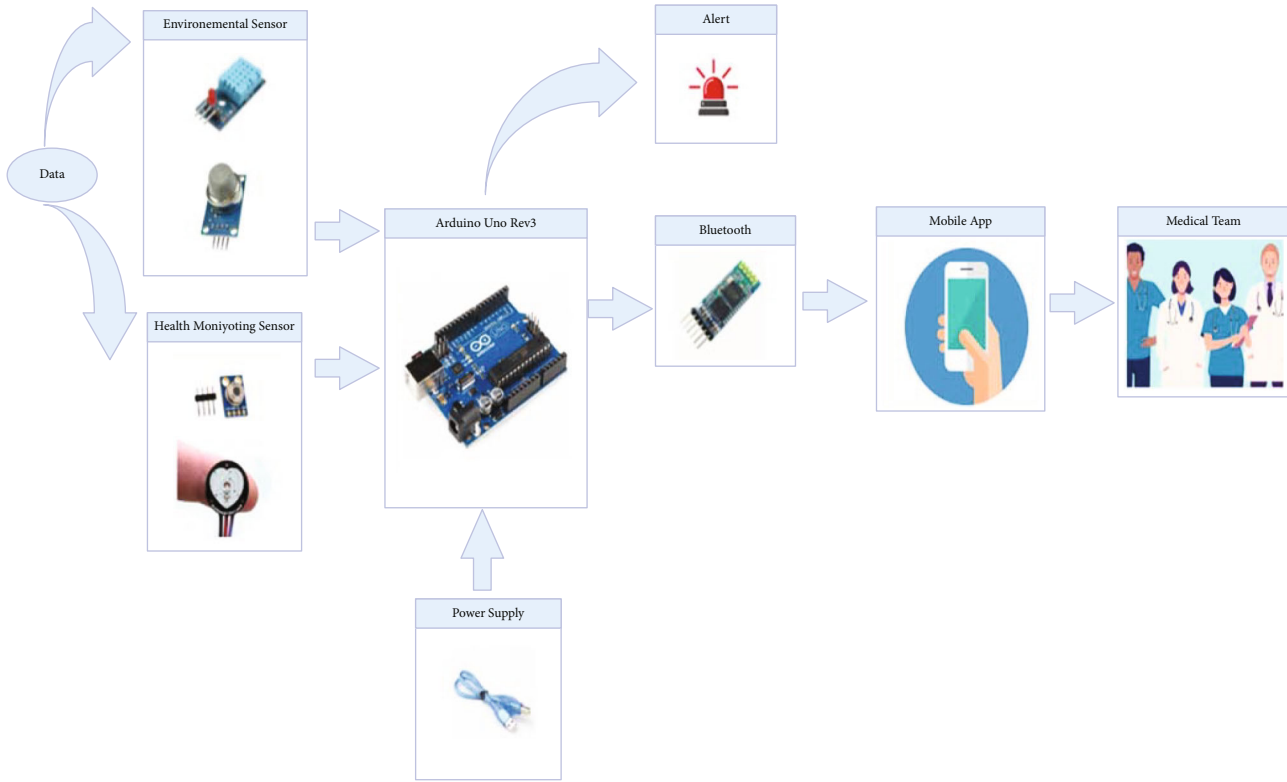

Figure 1: System architecture.

The novelty of the developed system is that, unlike earlier research systems discussed in this paper, it monitors both environmental and health parameters, which are essential since asthma attacks are sudden occurrences that are affected by external factors like humidity, air quality, and ambient temperature. As asthma patients are at the utmost risk of contracting COVID and deteriorating their health further, they should be kept under constant supervision and care. This portable system and the mobile application benefit us by taking real-time data and immediately alerting officials if there is any abnormality in the readings. When the level of a parameter falls or exceeds the required limits, the real-time monitoring solution can generate warnings. Thresholds and tolerances can be set for each sensor so that when the level fluctuates outside of a certain range, notifications are sent right away. These alerts are available through a variety of methods, including push notifications, SMS, email alerts, and even in-app messages. This makes it incredibly easy to keep track of a patient's health over a set period of time. This makes the system more user-friendly than any other previous system, as the user can also track the state of the surroundings and body conditions of the patients even without the help of any doctors or medical professionals. In this system, for example, if the heart rate goes beyond the normal boundary, it is a possible sign of an asthma attack; hence, the system will instantly alert both the user and the medical staff through emergency signals like the onset of a buzzer and messages. This way, transportation costs are reduced for the patients, and doctors can make urgent decisions while maintaining a physical distance during the pandemic.

In the following section, the paper discusses the methodology extensively, including the functionalities of the various hardware modules used for the purpose of this system, the mobile application, the cost table, the working flowchart, and the circuit diagram. The outcomes of the system are discussed in Section 3, where each parameter is analyzed elaborately through appropriate tables and line charts. Section 4 provides a summary of the entire research where it emphasizes the necessity and efficiency of the system and how it can be improved for future scope.

## 2. Methodology

*2.1. Methodology Statement.* This section focuses on the methodology, the components, and the necessary steps to achieve the system's goal. As shown in Figure 1, the IoT-based health monitoring system is divided into three phases following the three-layer IoT architecture—data sensing, data processing, and data visualization. At the heart of the system is the Arduino Uno board, with its microcontroller, which processes the digital or analog data collected and transmitted to it by the sensors. The microcontroller then sends this data to the serial monitor of the Arduino IDE and mobile application via Bluetooth, which acts as the gateway.

The sensors will do the data sensing job in the system as shown in Figure 1. They will be connected with the bodies of the patients to collect real-time data. That data is then converted to a human-readable version by the Arduino UNO and displayed in the mobile application.

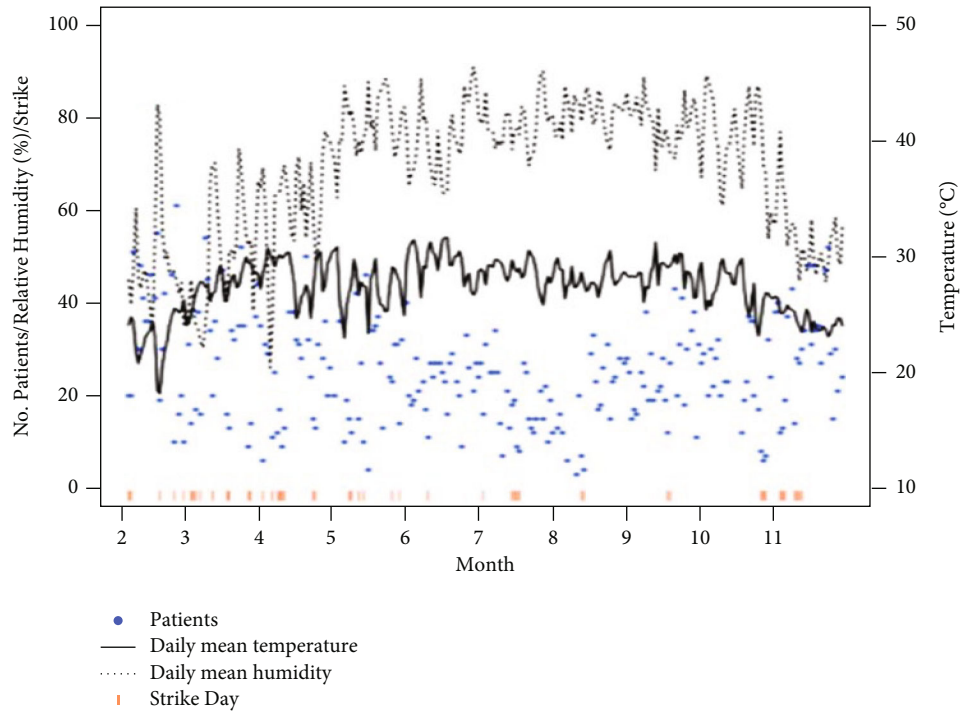

Figure 2: Effect of ambient temperature on daily nebulized asthma hospital visits in a tropical city of Dhaka, Bangladesh.

### 2.2. Modules and Materials

#### 2.2.1. Arduino UNO.
The Arduino UNO is the most crucial component of the system. It is necessary for interconnectivity and data transmission or reception, which are fundamental to the IoT. This microcontroller works as the heart of the system, which will convert the data received from the sensors and pass it to the mobile application.

There are six analog pins that receive analog signals and convert them to digital data, which is utilized for the Arduino's connection with analog sensors connected to the body of the patients. The board also has 3.3 V and 5 V output voltage pins which can be used to power the Arduino components, and more than one ground pin. The Arduino board has TX (1) and RX (0) pins for establishing serial communication in the system, mainly through the external Bluetooth module. Data transmission occurs through the TX pin, and reception occurs through the RX pin, both of which are indicated by the corresponding flashing of TX and RX LEDs on the board. The data can be sent or viewed through the serial monitor of the Arduino IDE.

#### 2.2.2. Sensors.
A total of four sensors have been used for the purpose of this system. The health monitoring sensors MLX90614 and pulse sensor are wearable devices attached to the patient's body to fetch data through physiological signals like body temperature and heart rate. The environmental monitoring sensors DHT-11 and MQ-135 detect the condition of the humidity and air quality, respectively. The analog signals received by the Arduino from the sensors are converted to digital data.

*(1) Infrared Temperature Sensor—MLX90614.* One of the parameters to be measured in this system is temperature. Fluctuations in the ambient temperature, which is the temperature of the patient's surroundings, can cause asthma to become severe in patients of different age groups. In the warm climate of Bangladesh, it is observed that a decrease in ambient temperature is correlated with more asthma patients being admitted to the hospital, particularly in adults. This is mainly because drier air triggers breathing issues and more mucus secretion, indicating inflammation in the lungs. It is necessary to maintain and monitor a mild ambient temperature to prevent asthma attacks. The graph in Figure 2 [18], which was obtained during a study conducted in Dhaka, Bangladesh, in 2013, shows that the number of patients hospitalized is significantly lower at temperatures around 27.8 degrees Celsius.

Since COVID-19 is a respiratory illness, it puts asthma patients at a greater risk of being hospitalized. Due to this, it is also necessary to monitor the body temperature of the patient to detect if they are running a fever. Early detection of COVID symptoms like fever can help prevent asthma from worsening.

Both the ambient and body temperatures can be measured in degrees Celsius through the infrared temperature sensor (MLX90614) contained in the system. This sensor is able to measure both the ambient and body temperature without contact, based on the amount of infrared radiation emitted. The sensor is connected with the Arduino board and can pass data from the environment and also from the body of the patient so that doctors or medical personnel can detect the issues regarding the condition immediately.

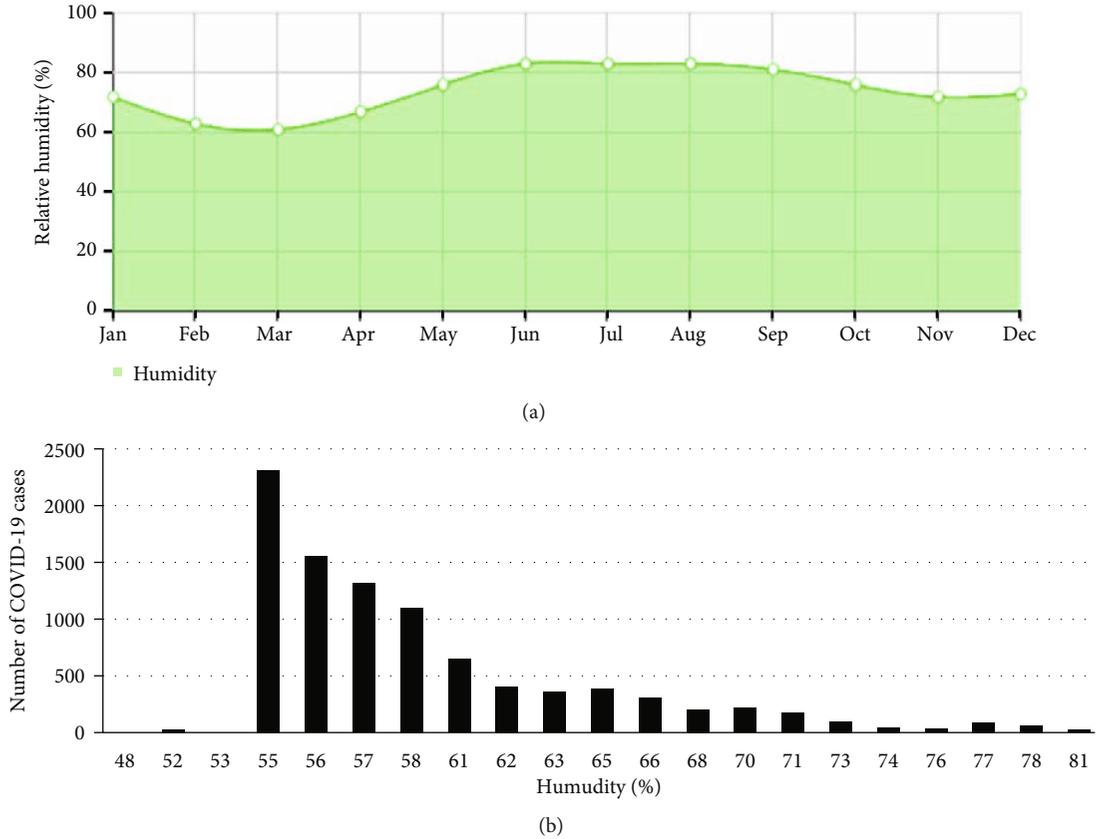

Figure 3: (a) The mean monthly relative humidity over the year in Dhaka, Bangladesh. (b) Cases of the COVID-19 by the amount of humidity (%), in Bangladesh from March 08 to May 03, 2020.

*(2) Humidity Sensor—DHT11.* Another parameter measured by the system is humidity. Bangladesh is a highly humid country with an average of 74.0% annual humidity. When the air is humid, it becomes difficult to breathe because the airways in the lungs become constricted, and this is particularly detrimental for asthma patients. However, it has been observed that higher humidity levels are associated with a lower number of COVID-19 cases.

An ideal indoor humidity level of at least 60% should be maintained to control the possibility of an asthma attack and to lessen the chances of coronavirus transmission. Figure 3 (a) [19] shows the variation in humidity percentage throughout 2021 in Bangladesh, with the peak being in August. Figure 3(b) [20] depicts the fall in COVID-19 cases with rising humidity.

To measure humidity levels, a sensor called DHT-11 has been used in the system. It consists of a capacitor whose resistance changes according to how humid the air is. It also has a built-in integrated circuit that converts the incoming analog data to digital data. The output pin of the sensor is connected to any digital pin of the Arduino Uno.

*(3) Air Quality Sensor—MQ135.* The system also measures the air quality of the environment. Pollutants found in the

Table 1: Categories of air quality index.

| AIR quality index (AQI) | Category |
|---|---|
| 0–50 | Good |
| 51-100 | Satisfactory |
| 101-200 | Moderate |
| 201-300 | Poor |
| 301-400 | Very poor |
| 401-500 | Severe |

air can severely affect our respiratory systems and trigger asthma. These air contaminants consist of ground-level ozone gas and other harmful particles such as dust, soot, chemical vapors, and dissolved gases. According to the World Air Quality Report 2020, Bangladesh has been named the country with the worst air quality index in the world. Table 1 [21] shows the level of impact of different ranges of the air quality index. As claimed by the Eco-Social Development Organization (ESDO), the number of asthma patients rose to 78,806 from 3,326 and deaths due to asthma increased ten times, from 56 to 588, in the span of only 5 years, from 2015 to 2019.

Research studies have shown that air pollution reduces immune responses in humans, enabling viruses such as the coronavirus to become severe in most cases. The rising trend

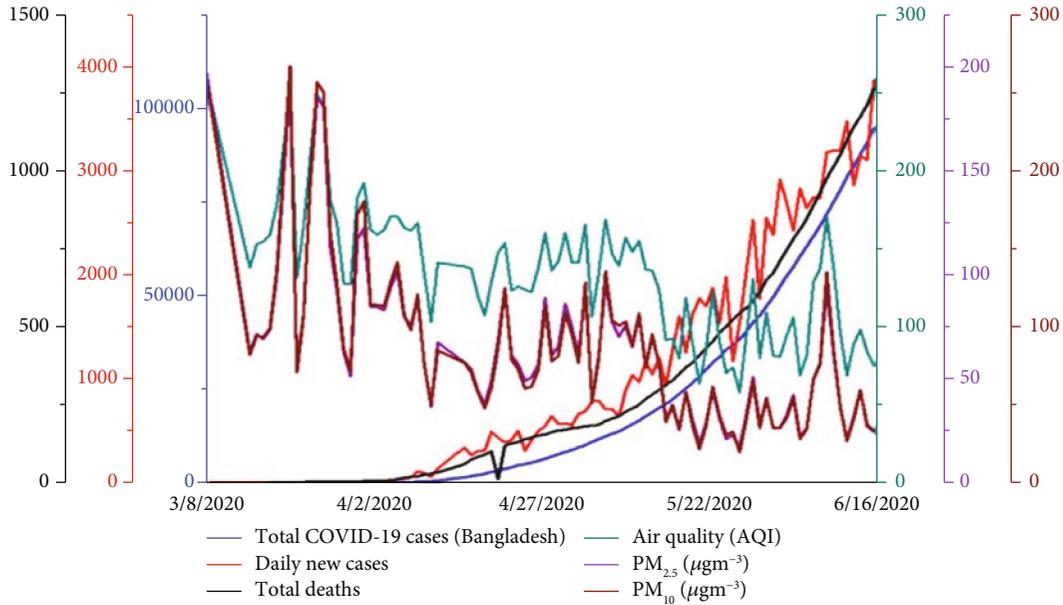

Figure 4: Depiction of COVID-19 cases and air quality (PM2.5, PM10, and AQI) from March 8 to June 16, 2020, in Bangladesh.

Table 2: Average heart rate in people from age 0 to above 16.

| Age | Range |
| --- | --- |
| 0-1 month | 100-180 |
| 2–3 month | 110-180 |
| 4–12 month | 80-180 |
| 1–3 years | 80-160 |
| 4–5 years | 80-120 |
| 6–8 years | 70-115 |
| 9–11 years | 60-110 |
| 12–16 years | 60–110 |
| >16 years | 60-100 |

Table 3: Cost table for the developed system.

| Components | Price in BDT | Approximate price in USD |
| --- | --- | --- |
| Arduino Uno R3 board | BDT 599 | USD 7.00 |
| Gas sensor MQ135 | BDT 120 | USD 1.40 |
| Humidity sensor DHT 11 | BDT 240 | USD 2.81 |
| Temperature sensor MLX90614 | BDT 1099 | USD 12.84 |
| Pulse sensor | BDT 290 | USD 3.39 |
| Bluetooth module HC 05 | BDT 400 | USD 4.68 |
| LCD display | BDT 310 | USD 3.62 |
| Breadboards x 2 | BDT 102/pc | USD 1.19/pc |
| Jumper wires | BDT 130/box | USD 1.53/box |
| Led lights | BDT 6/pc | USD 0.70/pc |
| Buzzers | BDT 15/pc | USD 0.18/pc |
| Total | BDT 3440 | USD 40.12 |

in both the daily new COVID cases and the air quality index can be observed in Figure 4 [22].

A gas sensor MQ-135 has been used in the system to measure the air quality of the surroundings. This sensor utilizes tin (IV) oxide, which has a higher resistance than air. This sensor utilizes tin (IV) oxide, which has a higher resistance than air. The concentration of gases and the resistance of $SnO_2$ are inversely correlated. The resistance of the sensor in the presence of gases is calculated from the resistance of the sensor when the air is clean. As the sensor provides data to the Arduino, the output can be achieved in ppm through any of the analog pins of the Arduino Board.

*(4) Pulse Sensor.* This system can also measure the number of heartbeats in a span of 1 minute. Table 2 [23] shows that the normal heart rate for an average adult is 60-100 bpm. The rate at which the heart beats is a factor in considering how healthy a person is. In asthmatic patients, a varying heart rate is quite common due to the chemoreceptor reflex stimulated by hypoxia. The COVID symptom study app conducted a survey worldwide and came to the conclusion that the heart rate of a person can aid in finding out if a person has contracted the virus. These researchers have also deduced that an increased heart rate, possibly over 100 bpm, is a significant symptom of the coronavirus.

For this system, a regular pulse sensor has been used to measure the heart rate in bpm. The sensor contains a light that measures the pulse rate. Upon placing the patient's finger on the sensor, the light changes depending on how much blood has been pumped into someone's capillaries. The output is the result of the variation of light in the sensor.

*2.2.3. Bluetooth HC-05.* In this system, wireless data transmission and reception with the mobile application developed for this system occur through a Bluetooth HC-05 module. This Bluetooth module has two modes of operation: command mode and data mode. Commands are used in order to change the settings and parameters of the module, as well as to decide whether it will act as the master or slave in a communication pair. Data mode is used when data is being transferred to another device supporting Bluetooth. This module helps the user see the data in the mobile application that has been converted on the Arduino board.

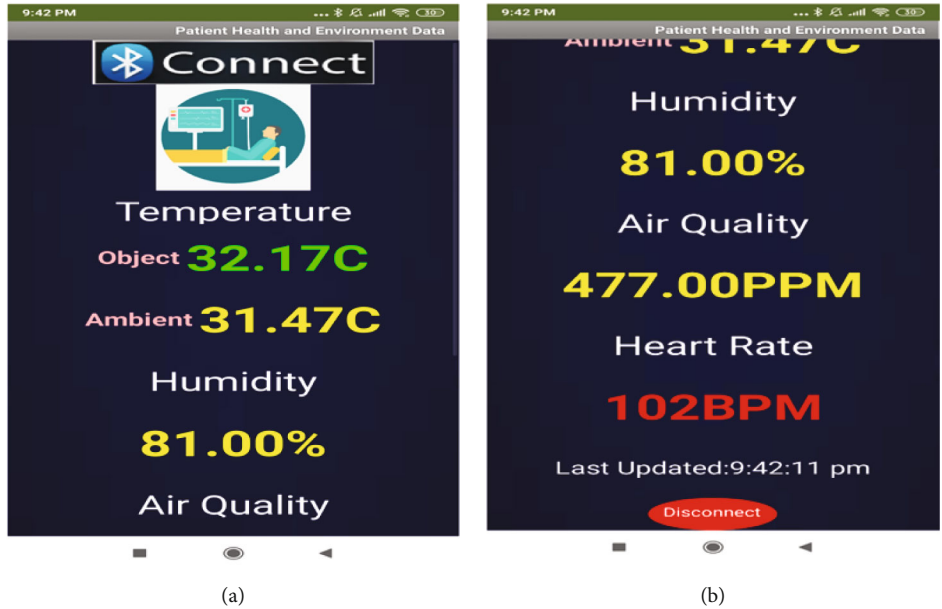

Figure 5: Screenshots of the mobile app—IoT healthcare.

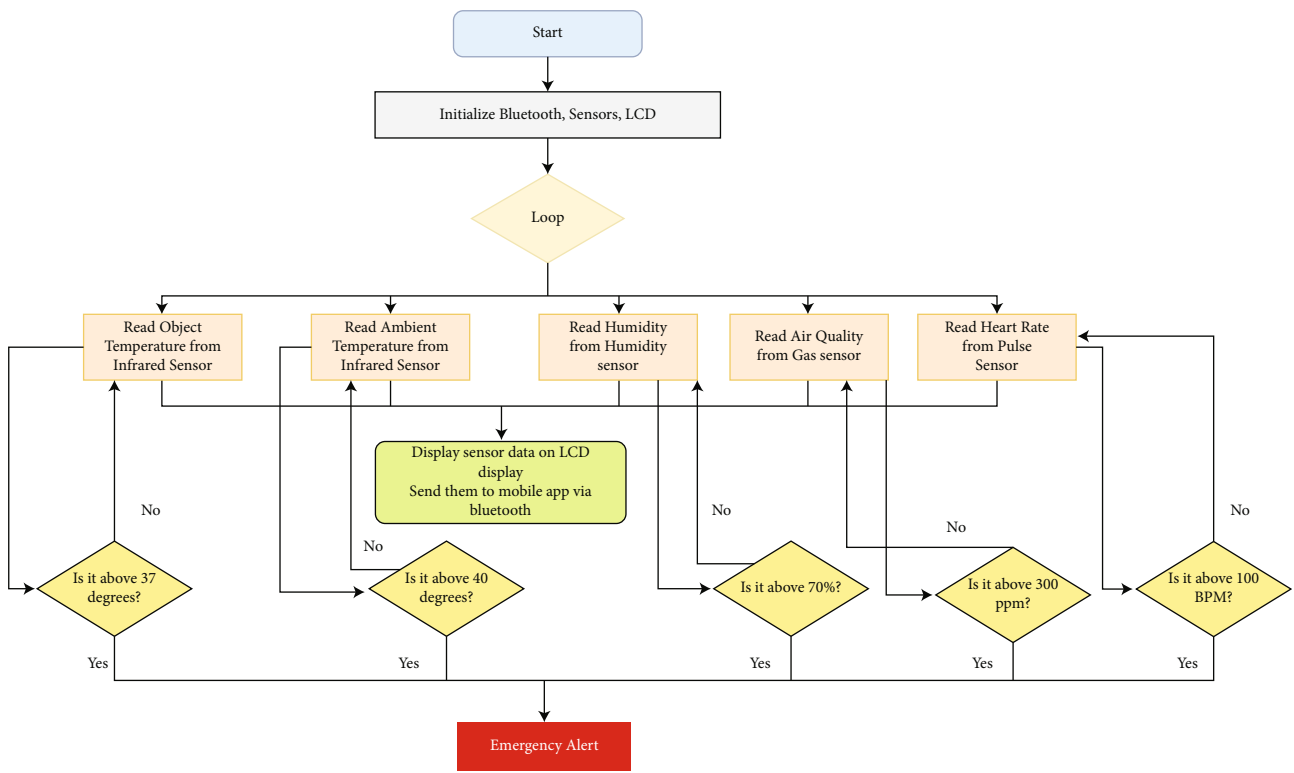

Figure 6: Working process flowchart.

*2.2.4. Cost of the Proposed System.* The sensors mentioned in Table 3 are able to receive data from one individual and their surroundings at a time. The total cost estimation of this project is approximately BDT 3440 or USD 40 only. Remote patient monitoring systems usually cost approximately USD 1600 per bed over a year [24], compared to which our system is extremely cost-efficient for people from all walks of life.

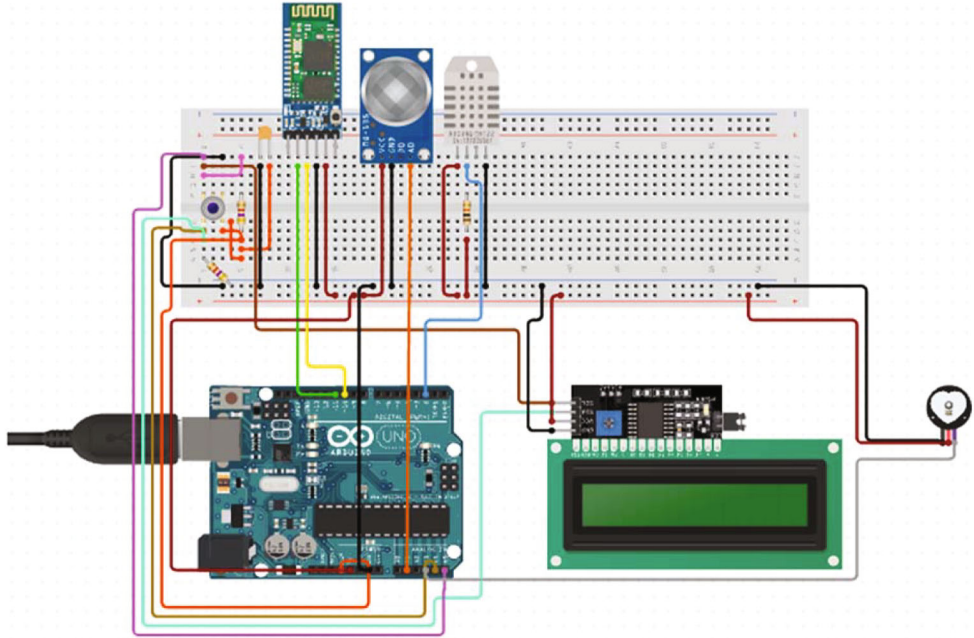

Figure 7: Circuit diagram of the developed system.

*2.2.5. MIT App Inventor 2.* For the monitoring purpose of this system from the medical staff's side, an Android app called "IoT healthcare" has been created using MIT App Inventor 2. MIT app inventor is an integrated development environment that makes it easier to build Android applications using blocks of prewritten code. The user interface is very easy to design by placing the necessary components on the simulated mobile screen. The application created by the MIT app inventor can be test-ran on mobile with the help of the MIT AI2 Companion. There is also the option of downloading the app on a mobile phone.

IoT healthcare connects to the system and receives data through the HC-05 Bluetooth module. Once connected, the app starts showing the live parameters of the patient and their environment. As shown in Figures 5(a) and 5(b), the parameters change color in accordance with the severity of the patient's conditions. For example, the color green indicates that the values are within the normal threshold.

The color yellow indicates that the values are moderate. When the color changes to red, it means that the parameters have reached an emergency level and the patient requires immediate attention from the medical team. This alerts and allows the doctor to make quick decisions regarding the treatment of the patient. They can also view when the values were last updated when they scroll down to the bottom of the screen.

*2.2.6. Full System Review and Working Process Flowchart.* The system operates according to the sequence of the instructions written in the Arduino program using the IDE. The program consists of several variables, functions, and libraries for the LCD, sensors, alarms, and Bluetooth module. It also consists of decision-making blocks for emergency situations. The sequence is shown in the working flowchart in Figure 6.

A working flowchart makes it easier to view the sequence in which each execution takes place. First, the Bluetooth module, LCD display, and all the sensors are initialized. Then, a process runs in a loop where each sensor reads the incoming data from the patient and the environment they are in. The data is displayed on the LCD, the serial monitor of the Arduino, and sent to the app through Bluetooth. A number of if-else conditions are checked to see if any of the values are abnormal, in which case an emergency alert goes off.

*2.2.7. Circuit Diagram.* The circuit diagram in Figure 7 for the system has been generated using the software called Circuit.io, an online tool for designing electrical circuits involving Arduino. The software automatically fixes all the components together when the building blocks are added by the user. The hardware implementation of the developed

Table 4: List of participants.

| Health status | Participants | Age range | Gender |
|---|---|---|---|
| Normal | Person 1 | 18-25 | Female |
| Normal, allergic rhinitis | Person 2 | 18-25 | Female |
| Anemia | Person 3 | 40-50 | Female |
| High blood pressure | Person 4 | 40-50 | Male |
| Diabetes | Person 5 | 50+ | Male |

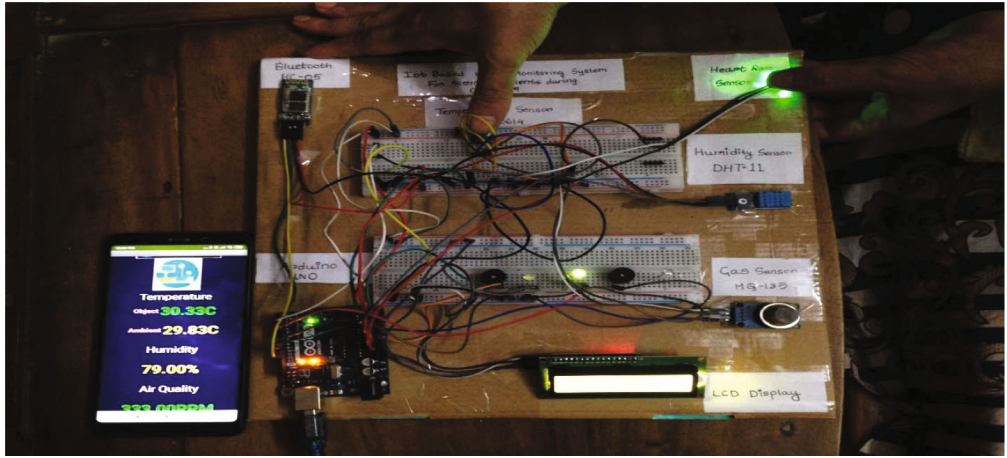

(a)

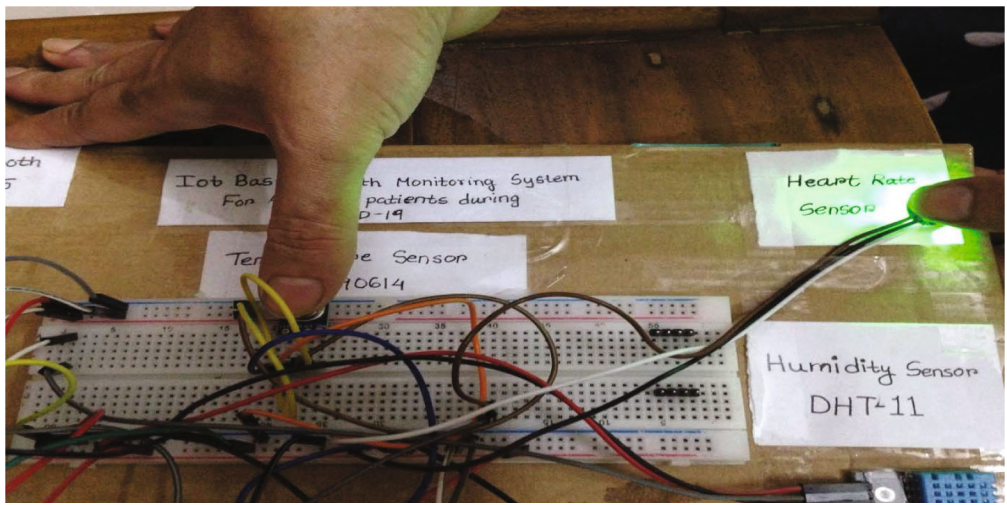

(b)

Figure 8: (a) Prototype for measuring patient's health and environment data. (b) Prototype while measuring patient's health and environment data.

system has a close resemblance to the schematic produced using the software.

In this circuit diagram, the sensors are connected to the Arduino Uno board through appropriate analog and digital pins and powered through the 5 V and 3.3 V ports accordingly. The LCD display is connected to the SDA and SCL terminals on the breadboard, which are in turn connected to the A4 and A5 pins of the Arduino, respectively, and powered through the 5 V port. The Bluetooth module is connected to the digital input/output pins for TX and RX communication.

## 3. Result Analysis

*3.1. List of Participants Examined.* The health monitoring system was tested on people of different ages and physical conditions over the span of one or two days at hourly intervals. All the participants were aged between 18 and 60 years. It should be noted that all the subjects participated in the test while they were completely rested. Deviations in body temperature and heart rate could be due to the movement of the participants' hands. Inaccuracy could also be a result of the movement of the sensors. Table 4 shows the list of people who took part in the test, specifying their age range, gender, and health conditions based on BMI and medical history.

*3.2. System Prototype.* When the system is turned on, the LCD screen shows that the system has initialized. Figure 8 (a) shows the prototype of the system worn by a patient. The temperature sensor gives two readings in degrees Celsius—the temperature of the patient and the ambient temperature. Since it is a noncontact sensor, the patient's body temperature can be measured from a small distance, as shown in Figure 8(b). If it detects an abnormal temperature in the patient or the environment, it triggers a buzzer and displays an emergency message on the LCD display and an emergency alert on the mobile app. The gas sensor reads the level of gases in ppm present in the room, and the humidity sensor measures the humidity in the room in percentage. If there is an abnormal reading, a white LED starts blinking; an emergency message is displayed both on the LCD and through the mobile app. The pulse sensor is touched by the patient with their finger and it reads their pulse rate in beats per minute. The patient should be careful about not touching the sensor with too much pressure to avoid inaccurate readings. An emergency message alert is sent in case the readings are lower or higher than a specified level.

Table 5: Body temperature data in °C.

| Hour | Person 1 | Person 2 | Person 3 | Person 4 | Person 5 |
| --- | --- | --- | --- | --- | --- |
| 1st | 34.19 | 33.33 | 31.17 | 36.50 | 33.96 |
| 2nd | 33.20 | 34.49 | 30.55 | 35.20 | 34.23 |
| 3rd | 34.63 | 34.81 | 31.04 | 37.89 | 33.60 |
| 4th | 34.43 | 31.00 | 31.64 | 35.95 | 33.02 |
| 5th | 34.01 | 33.00 | 32.23 | 36.52 | 32.88 |
| 6th | 34.12 | 32.62 | 31.96 | 37.54 | 33.97 |
| Average | 34.10 | 33.21 | 31.43 | 36.60 | 33.61 |

Table 6: Heart rate data in bpm.

| Hour | Person 1 | Person 2 | Person 3 | Person 4 | Person 5 |
| --- | --- | --- | --- | --- | --- |
| 1st | 68 | 81 | 77 | 82 | 102 |
| 2nd | 71 | 86 | 78 | 87 | 95 |
| 3rd | 74 | 83 | 72 | 92 | 99 |
| 4th | 65 | 95 | 71 | 85 | 92 |
| 5th | 77 | 83 | 65 | 96 | 88 |
| 6th | 72 | 80 | 73 | 88 | 91 |
| Average | 71 | 85 | 73 | 88 | 95 |

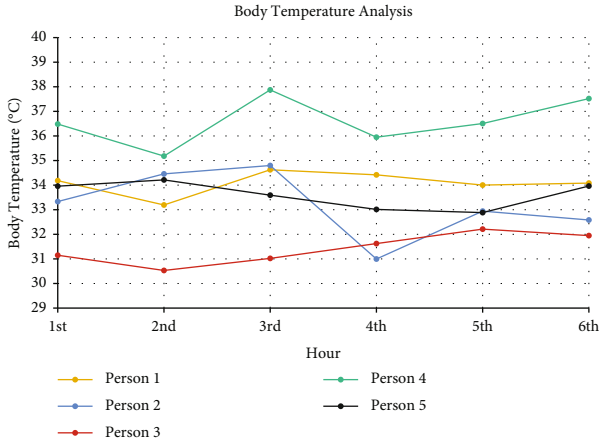

Figure 9: Line chart for body temperature data.

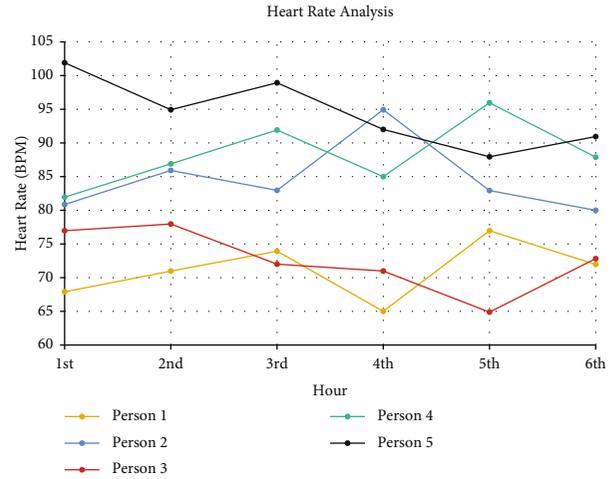

Figure 10: Line chart for heart rate data.

3.3. Data Sheet and Analytical Charts for Different Parameters. This section shows the elaborate analysis of the variation in each of the parameters for each participant when the system is used to monitor their health over a span of 6 hours.

3.3.1. Body Temperature Analysis. This system was used to measure the body temperature of five people over the course of six hours. Body temperature fluctuates according to factors like age, gender, health condition, and even the time of the day. Table 5 shows the datasheet of their body temperature that resulted from the monitoring process. Each participant was sat, and their data was continuously taken every second for an hour to see how high or low their body temperature reached throughout the specified period. Several rows of data were collected for each hour and then averaged to find the highest occurring temperature for that hour. A core body temperature of 38.1 degrees Celsius or higher is usually considered to be a fever.

From the average values in Table 5, we see that none of the participants suffer from fever, so we can conclude that all the participants are free from a major symptom of COVID-19 within the specified time period. Among all the participants, person 1, who is physically healthy, has the most stable body temperature, as their readings were consistent throughout the process. Person 2, who is in the same age range, had more variations in their values compared to person 1. Despite their age and underlying health condition, person 5 had a similar average body temperature to person 1 without any abnormalities. However, person 4 consistently showed relatively higher temperatures than everyone else, which could be linked to their condition of having hypertension. On the other hand, person 3 had the lowest average body temperature, which could be because of their anemia.

Figure 9 shows a comparison between the variations in body temperatures for all the participants according to their age, body status, and gender through a line chart. From the line chart, it is clear that person 4 consistently has the highest body temperature among all the participants. In addition to that, their data showed a huge deviation between the 2nd and 4th hour, where it reached close to 38 degrees Celsius. Person 2 had the most uneven fluctuations in their body temperatures. However, seeing the line chart, we can be sure that none of the participants, except for person 4, were at risk of developing a fever.

3.3.2. Heart Rate Analysis. The variations in the heart rate data of the five participants over the same interval are displayed in Table 6. As discussed earlier, an irregular heartbeat is associated with asthma, while an increased heart rate could indicate coronavirus infection. From the table, it can be seen that person 1 and person 3 show a relatively healthy range of heart rate over the six hours of examination. Person 2, person 4, and person 5 all have heart rates above 80 bpm which could be due to asthma, COVID-19, or their preexisting health conditions. Due to their medical history, it is difficult to make a solid diagnosis about whether they are particularly suffering from asthma or COVID-19 with the help of this system. This system is useful to monitor patients after their diagnosis has already been carried out.

Table 7: Ambient temperature data in °C.

| Hour | Person 1 | Person 2 | Person 3 | Person 4 | Person 5 |
|---|---|---|---|---|---|
| 1st | 31.17 | 31.32 | 30.37 | 30.95 | 30.23 |
| 2nd | 31.16 | 31.33 | 30.79 | 31.01 | 30.29 |
| 3rd | 31.12 | 31.43 | 31.16 | 30.96 | 30.30 |
| 4th | 30.10 | 29.39 | 30.26 | 30.99 | 29.66 |
| 5th | 30.04 | 29.54 | 30.32 | 31.08 | 29.85 |
| 6th | 30.11 | 29.65 | 30.30 | 31.08 | 29.96 |
| Average | 30.62 | 30.44 | 30.53 | 31.01 | 30.05 |

Table 8: Humidity data in %.

| Hour | Person 1 | Person 2 | Person 3 | Person 4 | Person 5 |
|---|---|---|---|---|---|
| 1st | 73.51 | 73.58 | 91.16 | 74.58 | 78.08 |
| 2nd | 73.65 | 73.51 | 90.79 | 73.99 | 78.01 |
| 3rd | 73.48 | 73.54 | 89.05 | 73.79 | 78.01 |
| 4th | 78.19 | 92.27 | 78.34 | 74.02 | 81.61 |
| 5th | 78.01 | 91.81 | 78.02 | 73.76 | 81.25 |
| 6th | 78.00 | 92.03 | 78.14 | 74.08 | 79.89 |
| Average | 75.81 | 82.79 | 84.25 | 74.04 | 79.48 |

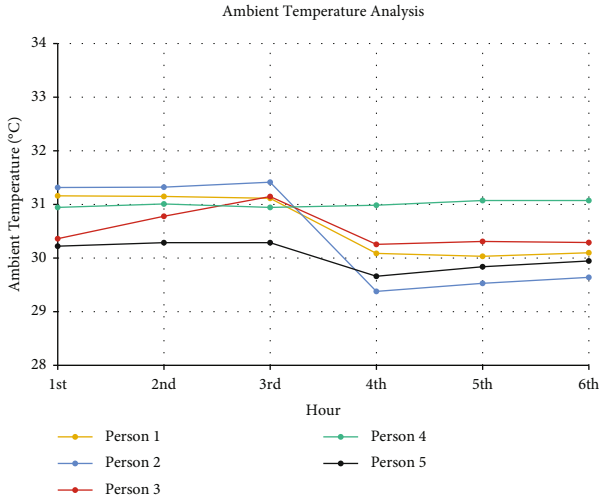

Figure 11: Line chart for ambient temperature data.

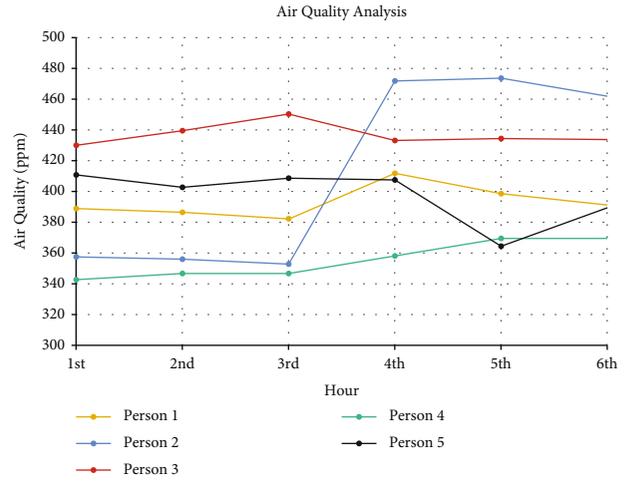

Figure 13: Line chart for air quality data.

Table 9: Air quality data in ppm.

| Hour | Person 1 | Person 2 | Person 3 | Person 4 | Person 5 |
|---|---|---|---|---|---|
| 1st | 389.44 | 357.76 | 430.44 | 343.20 | 411.53 |
| 2nd | 387.16 | 356.29 | 439.98 | 346.78 | 403.13 |
| 3rd | 382.64 | 353.10 | 450.96 | 346.72 | 409.18 |
| 4th | 412.36 | 472.61 | 433.65 | 358.46 | 408.13 |
| 5th | 398.80 | 474.30 | 434.90 | 369.87 | 364.83 |
| 6th | 391.67 | 462.15 | 434.34 | 369.68 | 390.67 |
| Average | 393.68 | 412.70 | 437.38 | 355.79 | 397.91 |

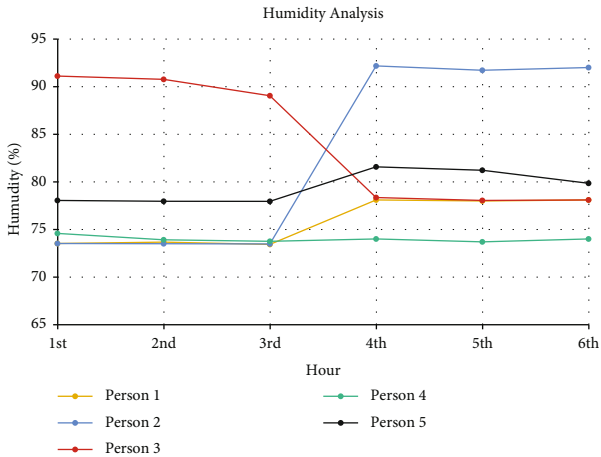

Figure 12: Line chart for humidity data.

Person 2 showed the most irregular pattern of heart rate among the group, with the largest deviation in the 4th hour. Their heart rate increased to 95 bpm from 83 bpm and then fell to 83 bpm again between the 3rd and 5th hour. This abnormal irregularity in heart rate strongly suggests the existence of asthma in the person. It should be noted that such an increase could also be influenced by their allergic rhinitis. Persons 4 and 5 have the highest average values for heart rate. Person 4 has high blood pressure, known as hypertension, which has a very strong link with an elevated heart rate known as tachycardia. High blood sugar, known as hyperglycemia, could also cause a rapid heart rate, as seen in the case of person 5.

Table 10: Comparison with other researches.

| Reference | Parameters | Result |
| --- | --- | --- |
| Paper [25] | Heart rate | 74.47 to 91.39 bpm<br>The heart rates of five people of different ages are in this range. |
| Paper [26] | Body temperature | 36.67 to 37.11°C<br>The body temperatures of twenty people aged 18–24 are in this range. |
| Paper [27] | Ambient temperature | 29.30 to 29.50°C<br>The ambient temperature varied at midnight in this range. |
| Paper [28] | Humidity | 68.89%<br>Grand average of the humidity data. |
| Paper [29] | Lung condition/vesicular breath sound | This research used an innovative approach of monitoring the vesicular breath sound of the patient using a modified stethoscope since COVID-19 is detrimental to the lungs. |
| This paper | Heart rate, body temperature, ambient temperature, humidity, and air quality | Heart rate: 71 to 95 bpm<br>Body temperature: 31.43 to 36.60°C<br>Ambient temperature: 30.05 to 31.01°C<br>Humidity: 74.04 to 84.25%, with a total mean of 79.27%<br>Air quality: 355.79 to 437.38 ppm. |

As shown in Figure 10, a line chart has been generated to display how the heart rate varied from one person to another over the six hours. There is some irregularity in everyone's data; however, it can be seen from the chart that person 5 consistently had a resting heart rate above 90 bpm, reaching as far as 102 bpm, which can have very serious consequences if not given proper medical attention.

3.3.3. Ambient Temperature Analysis. If a person has been diagnosed with asthma, it is particularly important to monitor the ambient temperature continuously. This was done with all five participants regardless of whether they were officially diagnosed with asthma or not. From the average readings in Table 7 for each person, we can see that there are little to no differences in the temperatures of their locations over the time period.

As shown in Figure 11, we used a line chart to analyze how changing ambient temperatures affect everyone individually. From the chart, it can be seen that there is little variation over the six hours of the monitoring process. However, strong fluctuations in the temperatures are noticed between the 3rd and 4th hours in the case of almost everyone. This could be due to the season, since all the people who participated in the test were tested at the same time of the month.

The chart shows that the ambient temperature was well maintained for person 4 the entire time. From the previous analysis of heart rate, it was observed that person 2 was possibly asthmatic. In that case, there was a very large drop in temperature in their location between the 3rd and 4th hours. Lower temperatures are detrimental to asthma and can lead to asthma attacks, so this is concerning for person 2. Significant drops in temperatures were also noticed in the locations of person 1, person 3, and person 5.

3.3.4. Humidity Analysis. While high humidity eliminates the risk of a higher rate of COVID infections, it raises the risk of asthma attacks by a significant amount. The line chart in Figure 12 depicts how the humidity levels of the locations of the five participants varied over the given hour. Figure 12 shows that, like the ambient temperature, the humidity levels also showed noticeable fluctuations between the 3rd and 4th hours for most of the participants. The average values of the humidity level can be observed from Table 8, which shows that person 2 and person 3's locations had readings above 80%, putting them at an increased risk of an asthma attack.

The humidity in person 4's location was the steadiest, which is important to reduce the chances of asthma attacks. Person 2's location had the largest jump in humidity, where it reached beyond 90%. This is dangerous since person 2 is highly likely to be affected by asthma. On the other hand, person 3's location showed the largest drop in humidity level, where it came down to below 80% and maintained the same humidity as that in person 1's location. The humidity levels in person 1 and person 5's locations showed relatively smaller changes. However, it should be noted that none of the humidity levels were within the safe threshold for asthma patients as they were all above 70%.

3.3.5. Air Quality Analysis. The changes in the air quality of each of the participants' locations over the time period can be seen in Figure 13. The average values in Table 9 indicate that most of the participants were in an environment where the air quality index was greater or equal to 400 ppm. An air quality index of over 400 ppm is considered extremely unhealthy and can raise the risk of asthma attacks and coronavirus infections significantly. It can be seen from Figure 13 that, in everyone's case, the air quality is not favorable at all, ranging from very poor to severe according to the categories shown in Table 1.

It is observed that in person 2's case, the spike in air quality index was the worst, as it reached beyond 460 ppm by the 4th hour and consistently remained so. Only the air quality in person 4's location was below 400 ppm; however, person 5 and person 1's locations also had a drop to lower levels after the 4th and 5th hours had passed, respectively. On the other hand, person 3's location had an air quality that consistently remained above 420 ppm. The preexisting medical conditions of person 2, person 3, person 4, and person 5 make them very susceptible to coronavirus infection because such poor air quality can weaken the immune system even further for these people.

*3.4. Comparative Analysis.* Many pieces of research are being conducted on IoT-based health monitoring systems, especially during the pandemic. Table 10 shows some of the similar health and environmental parameters that were mainly analyzed in such research, including some novel approaches taken by those research works concerning the similar issues addressed in this paper.

None of the previous research papers used for comparison with this study focused on asthma patients, particularly during the pandemic. Moreover, those systems were exclusively developed for monitoring health data or environmental data using IoT, but not both. Our system combines all the parameters concerning the patient's health as well as their surroundings. This is very crucial as asthma has a very strong linkage with not only a person's physical condition but also the condition of where they are located.

## 4. Conclusion

This is an IoT-based study that depicts how asthmatic patients during the COVID-19 pandemic can be monitored remotely. To execute this system, we used a microcontroller and some major health-specific sensors. A mobile app development application was used to build the mobile app. This research was conducted on five different people. The results were strong and well-founded in monitoring essential parameters from a distance and sending emergency alerts when needed. As shown in the results, person 2 had an extremely inconsistent heart rate, which is strong evidence of their asthma. However, it is difficult to deduce the cause of the irregularity in heart rate for others. Thus, this is primarily a system made not to detect diseases but to monitor them.

Currently, most people are under attack by asthma and its related issues. This could be due to the poor air quality of Bangladesh. From the given results, we see that the air quality has degraded significantly for most of our patients, which can be extremely dangerous for their immune systems. The increase in humidity levels and the drop in temperature could also make people more prone to viral infections. Hence, this system will be extremely beneficial for healthcare specialists to not only monitor patients but also monitor external factors. This will help both medical staff and patients, as it is more time-efficient and convenient since there has not been enough research conducted on asthma during the COVID-19 period. This system can grow into a sustainable project in the near future with the aid of advanced technology such as more high-tech sensors and GPS. An SPO2 sensor could be added to this system to better monitor a patient's oxygen saturation levels, which could prove to be helpful in being informed about their current health status. A GPS module can also be used to help track a patient's current location in case their symptoms deteriorate and they are in need of immediate help. This system can be further enhanced with the assistance of modern AI models, which can help detect more symptoms like coughing and shortness of breath and produce additional results to help monitor patients.

## Data Availability

No data was used to support the findings of this study.

## Conflicts of Interest

The authors declare that they have no conflicts of interest to report regarding the present study.

## Acknowledgments

The author would like to thank the Deanship of Scientific Research at Shaqra University for supporting this work.

## References

[1] "India: coronavirus cases," https://www.worldometers.info/coronavirus/country/india/.

[2] "Worldwide coronavirus cases," https://www.worldometers.info/coronavirus/.

[3] "Learn how to control asthma," https://cdc.gov/asthma/.

[4] "COVID-19 and asthma: what patients need to know," https://www.aaaai.org/Tools-for-the-Public/Conditions-Library/Asthma/covid-prevent.

[5] S. Anwar, M. Nasrullah, and M. J. Hosen, "COVID-19 and Bangladesh: challenges and how to address them," *Frontiers in Public Health*, vol. 8, p. 2, 2020.

[6] A. K. Mohiuddin, "An extensive review of patient health-care service satisfaction in Bangladesh," *Adesh University Journal of Medical Sciences & Research*, vol. 2, no. 1, pp. 5–16, 2020.

[7] Q. F. Hassan, "Core concepts: smart objects and smart environments," in *Internet of Things A to Z*, pp. 7-8, John Wiley & Sons, Inc., Chapter 1, Section 1.2, Hoboken, New Jersey, 2018.

[8] Y. Shelke and A. Sharma, *Internet of Medical Things: Benefits and Impacts of IoMT*, Aranca, Mumbai, 2016.

[9] L. Fernandes, "Internet of everything, Internet of medical things – the future of health IT," Internet of Everything, 2019. [Online].Available: file:///F:/paper/Paper%202022/Internet_of_Medical_Things_The_Future_of.pdf.

[10] B. Pradhan, S. Bhattacharyya, and K. Pal, "IoT-based applications in healthcare devices," *Journal of Healthcare Engineering*, vol. 2021, Article ID 6632599, 18 pages, 2021.

[11] J. T. Kelly, K. L. Campbell, E. Gong, and P. Scuffham, "The Internet of things: impact and implications for health care delivery," *Journal of Medical Internet Research*, vol. 22, no. 11, pp. 3-4, 2020.

[12] V. Tamilselvi, S. Sribalaji, P. Vigneshwaran, P. Vinu, and J. GeethaRamani, "IoT based health monitoring system," *International Conference on Advanced Computing and Communication Systems*, pp. 386–389, 2020.

[13] V. V. Gharbhapu and S. Gopalan, "IoT based low-cost single sensor node remote health monitoring system," in *7th International Conference on Current and Future Trends of Information and Communication Technologies in Healthcare*, vol. 113, pp. 408–415, Lund Sweden, 2017.

[14] N. A. Bassam, S. A. Hussain, A. A. Qaraghulli, J. Khan, E. P. Sumesh, and V. Lavanya, "IoT based wearable device to monitor the signs of quarantined remote patients of COVID-19," *Informatics in Medicine Unlocked*, vol. 24, pp. 100588–100614, 2021.

[15] S. R. Krishnan, S. C. Gupta, and T. Choudhary, "An IoT based health monitoring system," in *International Conference on Advances in Computing and Communication Engineering*, pp. 1–7, Paris, France, 2018.

[16] M. Hamim, S. Paul, S. I. Hoque, M. N. Rahman, and I. A. Baqee, "IoT based remote health monitoring system for patients and elderly people," in *2019 International Conference on Robotics, Electrical and Signal Processing Techniques*, pp. 533–538, Dhaka, Bangladesh, 2019.

[17] K. Lavanya, K. Nikitha, and S. R. Deepthi, "Low-cost wearable asthma monitoring system with a smart inhaler," in *2021 Asian Conference on Innovation in Technology (ASIANCON)*, pp. 1–7, PUNE, India, 2021.

[18] A. F. Kabir, C. F. S. Ng, S. Yasumoto, T. Hayashi, and C. Watanabe, "Effect of ambient temperature on daily nebulized asthma hospital visits in a tropical city of Dhaka, Bangladesh," *International journal of environmental research and public health*, vol. 18, no. 3, p. 890, 2021.

[19] "Average humidity in Dhaka," in *Weather & Climate*, Weather Spark, Dhaka, Bangladesh, 2021.

[20] S. E. Haque and M. Rahman, "Association between temperature, humidity, and COVID-19 outbreaks in Bangladesh," *Environmental Science & Policy*, vol. 114, pp. 253–255, 2020.

[21] A. Siddique, "Bangladesh grapples with polluted air," in *Dhaka Tribune: Special*, Dhaka Tribune, 2018.

[22] M. R. S. Pavel, M. R. Sarkar, A. Salam et al., "Impact and Correlation of Air Quality and Climate Variables with COVID-19 Morbidity and Mortality in Dhaka," vol. 5, no. 4, 2020https://www.medrxiv.org/content/10.1101/2020.09.12.20193086.abstract.

[23] W. M. Jubadi and S. F. A. M. Sahak, "Heartbeat monitoring alert via SMS," in *2009 IEEE Symposium on Industrial Electronics & Applications*, vol. 1, pp. 1–5, Kuala Lumpur, Malaysia, 2009.

[24] C. Mayer, *How Much Do Clinical Remote Monitoring Systems Cost?*, Medtronic, 2018.

[25] A. R. Dhruba, K. N. Alam, M. S. Khan, S. Bourouis, and M. M. Khan, "Development of an IoT based sleep apnea monitoring system for healthcare applications," *Computational and Mathematical Methods in Medicine*, vol. 2021, Article ID 7152576, 16 pages, 2021.

[26] S. Mishra, S. Shukla, S. Ravula, S. Chaudhary, and P. Ranjan, "Low-cost IoT based remote health monitoring system," *International Research Journal of Engineering and Technology*, vol. 6, no. 5, pp. 7984–7988, 2019.

[27] M. N. Hassan, F. Faisal, A. H. Siddique, and M. Hasan, "An IoT based environment monitoring system," in *3rd International Conference on Intelligent Sustainable Systems*, pp. 1119–1124, Thoothukudi, India, 2020.

[28] M. Z. Islam, M. A. Based, and M. M. Rahman, "IoT-based temperature and humidity real-time monitoring and reporting system for CoVid-19 pandemic period," *International Journal of Scientific Research and Engineering Development*, vol. 4, no. 1, pp. 1214–1221, 2021.

[29] S. S. Alam, M. S. Islam, M. M. F. Chowdhury, and T. Ahmed, "An IoT based health monitoring system to tackle COVID-19 in a contagious ward of hospital," *International Journal of Advanced Medical Sciences and Technology*, vol. 1, no. 4, pp. 11–18, 2021.